\documentclass[%
 reprint,
nofootinbib,
 amsmath,amssymb,
 aps,
]{revtex4-1}

\usepackage{graphicx}
\usepackage{dcolumn}
\usepackage{bm}
\usepackage{caption}
\usepackage{subcaption}
\usepackage{float}

\begin{document}


\title{Iterative Retraining of Quantum Spin Models using Recurrent Neural Networks}

\author{Christopher Roth}
\affiliation{
Physics Department, University of Texas at Austin
}

\date{\today}

\begin{abstract}
Modeling quantum many-body systems is enormously challenging due to the exponential scaling of Hilbert dimension with system size. Finding efficient compressions of the wavefunction is key to building scalable models. Here, we introduce iterative retraining, an approach for simulating bulk quantum systems that uses recurrent neural networks (RNNs). By mapping translations in the lattice vector to the time index of an RNN, we are able to efficiently capture the near translational invariance of large lattices. We show that we can use this symmetry mapping to simulate very large systems in one and two dimensions. We do so by 'growing' our model, iteratively retraining the same model on progressively larger lattices until edge effects become negligible. We argue that this scheme generalizes more naturally to higher dimensions than Density Matrix Renormalization Group.

\end{abstract}

\maketitle

\section{\label{sec:introduction}Introduction}

Many of the most interesting phenomena in solid state physics are driven by strong interactions, such as quantum hall physics \cite{quantumhall,quantumhallreview,anomhall} and unconventional superconductivity \cite{superconductivity,hightc}. Unfortunately, strongly interacting systems are also among the most difficult to simulate due to their extensive entanglement. This problem is exacerbated in two dimensions where edge effects can still be strong in systems with hundreds of electrons. 

As a result, simulating strongly interacting systems is as much a data science problem as a physics problem. For this reason, physicists have turned to machine learning for insights into modeling these enormous Hilbert spaces \cite{CarleoOrig,MLStrongCor,MLSignProblem,NAQS,LSTM,BosonsMachineLearning,BoltzmannReview,Tomography}. While machine learning models are more problem agnostic than physically informed Ansatz's (i.e. the Jastrow wavefunction \cite{jastrow}), they can overcome their lack of prior through greater expressivity. Furthermore, there is a massive infrastructure that allows machine learning models to take full advantage of modern computing capabilities. As an example, machine learning models are easily implemented on graphical processing units (GPUs) \cite{gpu}, which typically speed up computation by a factor of $10-100$.

In recent years, state-of-the-art performance has been achieved when important physical constraints are imposed on machine learning models. These constraints are often added post-hoc, through symmetrizing the output of the model \cite{NAQS} or telling the model the correct phase structure \cite{LSTM, j1j2}. In this manuscript, we discuss a symmetry of lattice systems that can be modeled directly by a machine learning architecture. We show that bulk translational invariance, an exact symmetry of infinite sized lattices and a nearly exact symmetry of macroscopic sized systems, can be imposed using an autoregressive model parameterized by a recurrent neural network. 

Recurrent neural networks (RNNs) are tremendously powerful tools that have have been used in language modeling \cite{RNNLanguage}, speech recognition \cite{speechRNN}, and image generation \cite{draw, pixelRNN}. RNNs process streams of data by maintaining a 'hidden state', which is updated by applying an identical function to the previous hidden state and the input at each timestep. As a result, the output of an RNN only depends on the relative timing of the input, not the global time index. This is important for sequence modeling, where context is dependent on relative displacements.

Similarly, electrons in solids communicate via the Coloumb interaction, which only depends on their relative distance. As a result, one can impose translational invariance by mapping the spatial component of the electron to the time component of an RNN. This is most naturally done on lattice models, where space is discretized, since RNNs typically have a discrete time index. 

We introduce a new simulation method, which we call iterative retraining, that takes advantage of this mapping between translationally invariant lattices and RNNs. Using an RNN based Ansatz, we are able to learn the ground state energy of extremely large lattices by continuously retraining the same model on progressively larger systems. We first learn a model of a small system, then use a small amount of data to generalize the model to a slightly larger system. We iterative over this process until we have trained a sufficiently large model to understand the bulk. Using iterative retraining, we are able to simulate extremely large systems in one and two dimensions by making efficient use of data.  

\begin{figure*}[t!]
\centering
  \includegraphics[width=.9\linewidth]{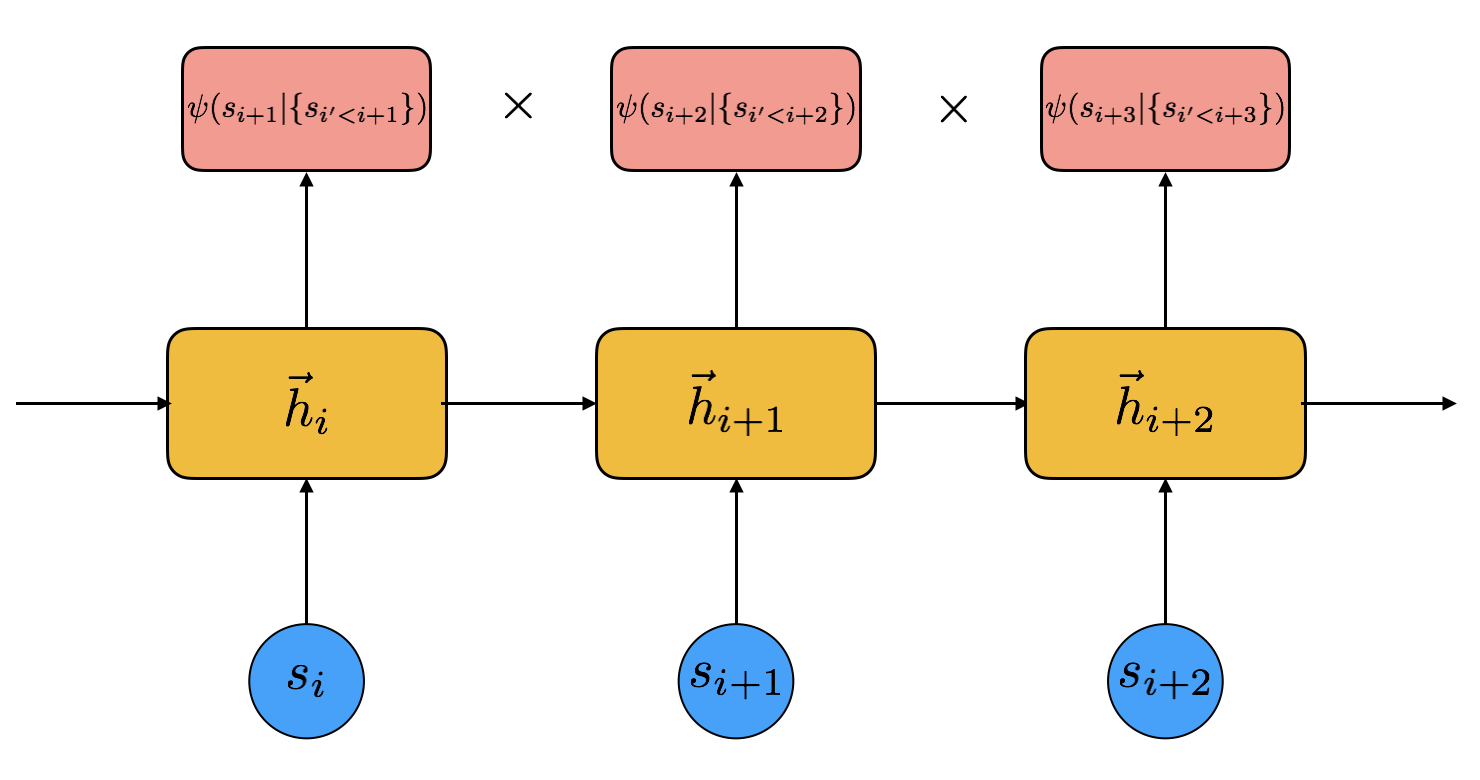}
  \caption{A recurrent neural network is driven by the spins along a 1D lattice and outputs the conditional wavefunction. The wavefunction for basis state $\vec{s}$ is the product of the conditional wavefunctions over all lattice sites.}
    \label{fig:1DLSTM}
\end{figure*}





\section{\label{sec:methods} Methods}

\subsection{\label{sec:VMC} Variational Monte Carlo}
In this paper we perform Variational Monte Carlo using an RNN-based Ansatz \cite{LSTM}. We learn the ground state wavefunction of a lattice of interacting spins by minimizing the energy of our wavefunction $\psi(\vec{s}, W)$ over variational parameters $W$, where $\vec{s}$ represents the z-component of the spins across the lattice. We compute the energy gradient $\frac{dE}{dW}$ stochastically by sampling over the norm squared of the wavefunction:
\begin{eqnarray}  \label{energy}
E = \sum_{\vec{s} \sim |\psi(\vec{s}) |^2 }  E^l_{\vec{s}},
\end{eqnarray}
\begin{eqnarray} \label{gradient}
\frac{dE}{dW} = \sum_{\vec{s} \sim  |\psi(\vec{s}) |^2} (E^{l*}_{\vec{s}} - E^*) \frac{d \textrm{log} ( \psi(\vec{s})) }{dW}, 
\end{eqnarray}
\begin{eqnarray}
E^l_{\vec{s}} = \sum_{\vec{s'}} \frac{H_{\vec{s}, \vec{s'}} \psi(\vec{s'})}{\psi(\vec{s})},
\end{eqnarray}
where $E^l$ is the local energy, $H_{\vec{s},\vec{s'}}$ are the matrix elements of the Hamiltonian connecting the basis states $\vec{s}$ and $\vec{s'}$, and $E$ is our stochastic estimate of the energy. We use the same sample over $|\psi{\vec{s}}|^2$ to compute both the energy and its gradient. 

We use this gradient information to make small updates to our parameters and minimize the energy. As is common in machine learning, we found that signal-to-noise based gradient averaging tended to work better than using exact gradients. All results in this manuscript use the Adam optimizer \cite{Adam}. 

\begin{figure*}[t!]
\centering
\begin{subfigure}{.45\textwidth}
  \centering
  \includegraphics[width=.9\linewidth]{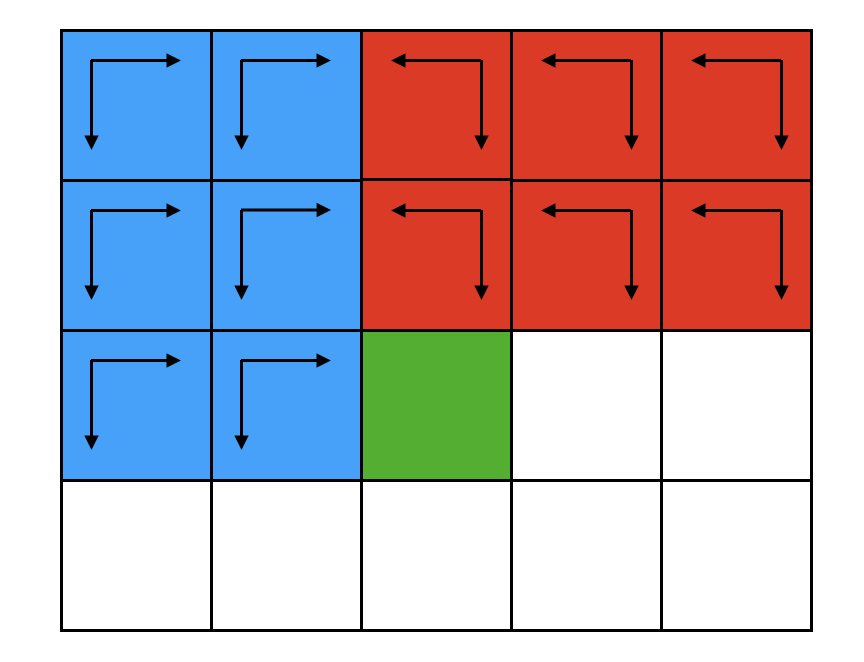}
  \caption{}
    \label{fig:context}
\end{subfigure}%
\begin{subfigure}{.45\textwidth}
  \centering
  \includegraphics[width=.9\linewidth]{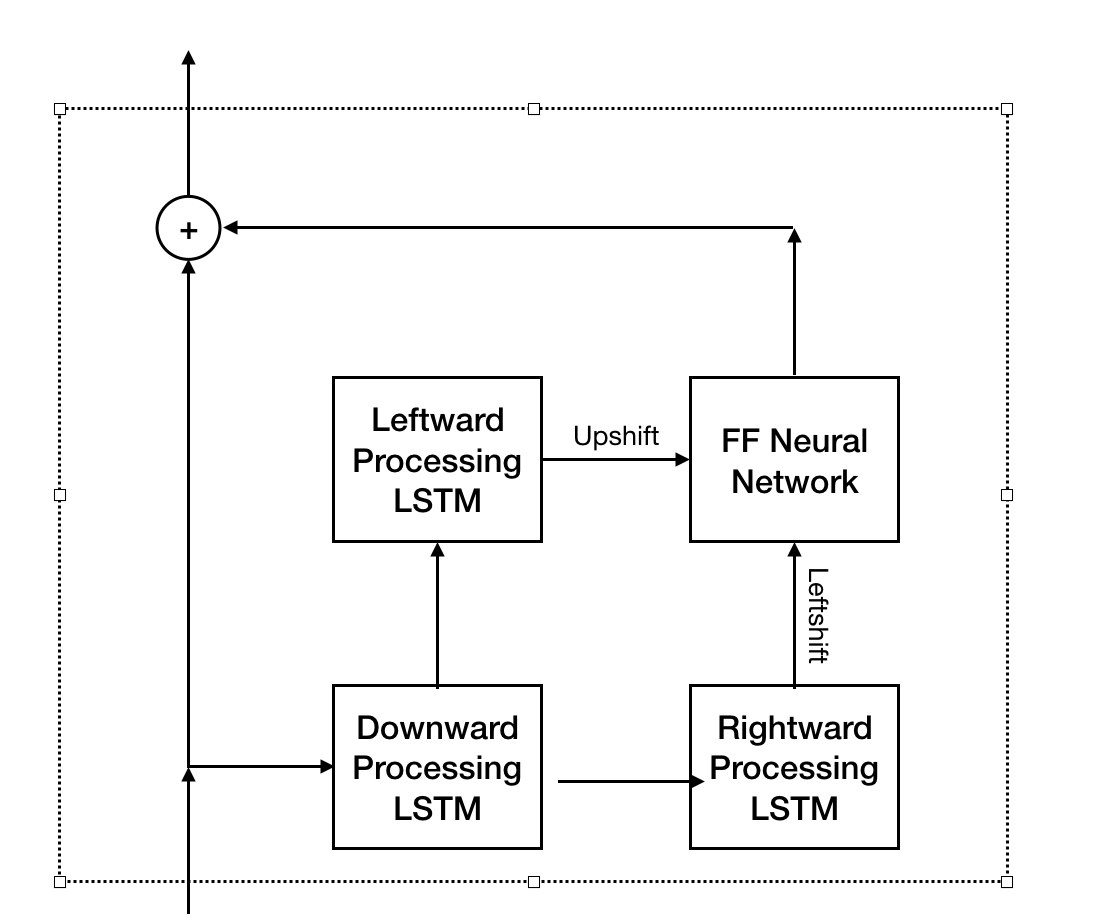}
  \caption{}
    \label{fig:architecture}
\end{subfigure}
\caption{a) Context in our model: In each layer we get context from two 'processors', one which integrates from top left to bottom right (blue) and the other which integrates from top right to bottom left (red). The outputs of processors are shifted left and up respectively to ensure the electrons are only conditioned on those before them. This context is translationally invariant for infinite sized systems. (b) The architecture of a single layer: We integrate over one direction at a time to improve parallelization.}
\end{figure*}

\subsection{\label{sec:autoreg} Autoregressive Ansatz}
We represent our wavefunction $\psi(\vec{s})$ using an artificial neural network, which takes a basis state $\vec{s}$ as input, and outputs its complex coefficient. We constrain our wavefunction to have the following form:
\begin{eqnarray}
\psi(\vec{s}) = \prod_i \psi(s_i|s_{i-1},s_{i-2},...s_1, \{W \}).
\end{eqnarray}
where $\psi(s_i|s_{i-1},s_{i-2},...s_1, \{W \})$ is known as the conditional wavefunction. The conditional wavefunction gives a probability amplitude and phase for each possible quantum state of each spin, contingent upon the quantum state of all spins behind it (based on an imposed ordering of the system). Spins are typically ordered moving from one side to the other, iterated along each dimension. This process, where the choice of spin at a particular site depends stochastically on the previous spins, is known as an autoregressive model. These models have been used to find the ground state wavefunction of many body systems, \cite{NAQS, RNNwf}, simulate classical spin models at finite temperature \cite{Tempwf}, and reconstruct entangled quantum states \cite{RNNreconstruct}. 

Factorizing the wavefunction in this way allows us to directly sample over $|\psi(\vec{s})|^2$ and stochastically estimate the expectation value of observables without Markov Chain Monte Carlo (MCMC) methods. As long as our conditional wavefunction is normalized, $\sum_{s_i} |\psi(s_i | s_{i-1}...s_1)|^2 = 1$, we can express the probability of a particular quantum state as a product over conditional probabilities:
\begin{eqnarray}
P(\vec{s}) = |\psi(\vec{s})|^2 = \prod_i |\psi(s_i | s_{i-1}...s_1)|^2.
\end{eqnarray}
A detailed proof is presented in \cite{NAQS}. Using this equality, we can sample over our the norm squared of our wavefunction by choosing our quantum numbers iteratively based on their conditional probabilities. These samples are independent and identically distributed, unlike those generated from MCMC methods \cite{NAQS}. 

In one dimension, we condition each particle on all particles to the left. In two dimensions, we condition each particle on all particles above, and particles to the left in the same row:
\begin{eqnarray}
\psi(\vec{s})= \prod_{i,j} \psi(s_{i,j} |\{ s_{i,j'<j}\}, \{s_{i'<i,j'}\}).
\end{eqnarray}

\subsection{\label{sec:model} Model}
We parameterize our autoregressive model using recurrent neural networks (RNNs), which encode information about sequences in a vector known as the 'hidden state'. The RNN applies an identical function at each time-step,
\begin{eqnarray}
\vec{h}_i = f(\vec{h}_{i-1}, s_i, \{W \} ), 
\end{eqnarray}
where $s_i$ is the spin at lattice site $i$, $\vec{h}_i$ is hidden state at site $i$, and \{W\} are the trainable parameters. In modern machine learning approaches, there are two common choices of functions: Long Short Term Memory networks (LSTMs) \cite{LSTM} and Gated Recurrent Units (GRUs) \cite{GRU}.  These architectures contain gates that regulate the flow of information and allow the network to maintain a persistant memory state \cite{GradientLSTM} (see appendix \ref{sec:LSTM} for more detail). We found that GRUs performed better in 1D, whereas LSTMs performed better in 2D. 

The conditional wavefunction is read out from the appropriate hidden state(s), so as to not break causality in the autoregressive model. In our model, we output a conditional probability that is normalized using the softmax function, and a conditional phase parametrized by the two vector $\{\sin{\phi}, \cos{\phi} \}$

In 1D, we use a single layer GRU and read out the conditional wavefunction from the previous hidden state as seen in figure \ref{fig:1DLSTM}. We use a two layer feedforward neural network for readout:
\begin{eqnarray}
\psi(s_i | s_{i-1}...s_1) = \textrm{feedforward}(\vec{h}_{i-1}, \{W \}).
\end{eqnarray}


In 2D, we employ a more complicated model, processing our input over a sequence of layers. Each layer consists of two 'processors', one which moves from the top left corner to the bottom right, the other which moves from top right corner to the bottom left. Using the output of both processors we are able to recover the full context for each electron as shown in figure \ref{fig:context}. Each processor consists of two LSTMs, the first of which is shared between the processors and integrates downwards along the sample,
\begin{eqnarray}
\vec{h}^{n,D}_{i,j} = f(\vec{h}^{n,D}_{i-1,j},\vec{h}^{n-1}_{i,j},\{W\}),
\end{eqnarray}
where $n$ indexes the layer, and $D$ indicates downward integration. Additionally, $\vec{h}^0_{i,j} = s_{i,j}$, since the first layer receives the lattice spins as input. The second LSTM takes input from the first LSTM and integrates to the right(left) for the corresponding processor:
\begin{eqnarray}
\vec{h}^{n,R(L)}_{i,j} = g(\vec{h}^{n,R(L)}_{i,j-(+)1},\vec{h}^{n,D}_{i,j},\{W\}).
\end{eqnarray}
We then shift these inputs left(up) and read out the next layer using 2-layer feed-forward neural network. We also include residual connections between the layers,
\begin{eqnarray}
\vec{h}^{n}_{i,j} = \vec{h}^{n-1}_{i,j} + \textrm{feedforward}(\vec{h}^{n,R}_{i,j-1}, \vec{h}^{n,L}_{i-1,j}, \{W\}).
\end{eqnarray}
The architecture of a single layer is shown in figure \ref{fig:architecture}. Since we only integrate along one direction at a time, the number of computation steps scales like the length of a side and we are better able to take advantage of the massive parallelism of GPUs. For the 2D results in this manuscript, we stack five of these layers and read out from a two layer feed-forward neural network. 

\begin{figure*}[t!]
\centering
\begin{subfigure}{.5\textwidth}
  \centering
  \includegraphics[width=.9\linewidth]{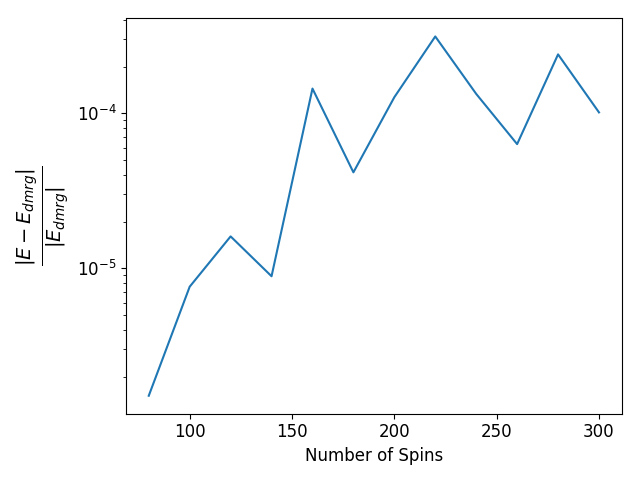}
  \caption{}  
   \label{fig:1dperformance}
\end{subfigure}%
\begin{subfigure}{.5\textwidth}
  \centering
  \includegraphics[width=.9\linewidth]{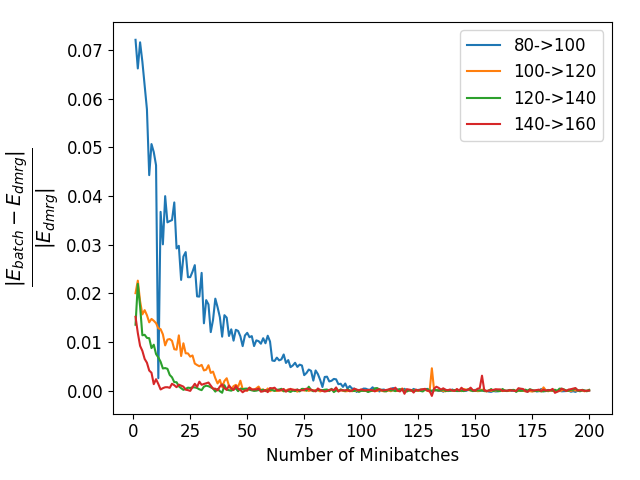}
 \caption{}
  \label{fig:iterative1d} 
\end{subfigure}
\caption{a) Error on the 1D Heisenberg model measured against DMRG as a function of number of spins: Energies of our trained models are estimated over 100,000 samples. With only $1000$ minibatches per iteration, we see good performance for systems up to 300 spins. (b) Generalization of the model scaling from N to N+20 spins: As the system gets larger, it takes progressively less samples to generalize.}
\end{figure*}

\subsection{\label{sec:iterativetraining} Iterative Retraining}

Since we use a recurrent neural network architecture, the shape of our model parameters is independent of system size. As a result, its possible to train the same architecture on systems of different lengths. For large systems that are close in size, we expect that the parameters for the ground state wavefunction are similar, as differences should only arise due to small edge effects. In light of this, we learn the ground state wavefunction for bulk systems by first training a small system, then iteratively retraining the same model progressively larger systems. We find that the amount of data needed to generalize decreases with system size. As a result, we are able to learn the ground state of extremely large systems by offloading the majority of training to smaller systems. 

\section{\label{sec:empiricalresults} Empirical Results}
We test our RNN Ansatz on the Heisenberg model in one and two dimensions. The Hamiltonian is given by,
\begin{eqnarray}
H = J_1 \sum_{<i,i'>} [s^x_i s^x_{i'} + s^y_i s^y_{i'} + s^z_i s^z_{i'}],
\end{eqnarray}
where $<i,i'>$ are pairs of nearest neighbor spins. Without loss of generality, we take $J_1 = 1$. The ground state is highly entangled, as the Hamiltonian contains raising and lowering terms in the $S_z$ basis. All of our results use open boundary conditions.
 
 \subsection{Results in 1D}
 
In 1D, we use iterative retraining to learn the ground state wavefunction for systems of up to 300 spins. We start by training a system of 80 spins and learn larger models by adding 20 spins at a time. We trained the 80 spin lattice for $5 x 10^5$ minibatches (10 GPU-hours), then generalized to larger lattices with only $1000$ minibatches at each lattice size. We compare our performance with density matrix renormalization group \cite{dmrg} calculations perfmored using the ITensor software \cite{ITensor}. We get good performance for systems up to $300$ spins as shown in figure \ref{fig:1dperformance}.
We found that our model did not need a lot of data to generalize; each time we added 20 spins, it took less than $100$ minibatches to have an error under $10^{-4}$ as shown in figure \ref{fig:iterative1d}. We also found that the number of minibatches needed to generalize decreased with system size.  

\subsection{Results in 2D}
Simulating the Heisenberg model in 2D is much substantially more difficult. We present two findings: 1) We show that an RNN based model can achieve low error on a small $6 \times 6$ system. 2) We use iterative retraining to scale up from a $6 \times 6$ system to a $30 \times 30$ system. Intriguingly, our error decreases as we scale up our system size, lending credence to the idea that our RNNs are learning to model the bulk. 

For our $6 \times 6$ model we use a five layer LSTM-based architecture as described in section \ref{sec:autoreg}. In order to maximize performance, we symmetrize our model as described in appendix $\ref{sec:appendix}$. The performance of our LSTM based model is shown in table \ref{table:2dheisenberg}, where we compare it with state-of-the-art variational Monte-Carlo \cite{PEPSheisenberg} and DMRG results \cite{dmrg6by6}.
\begin{table}[h]
\begin{center} 
\begin{tabular}{ |c|c|c| } 
 \hline
 PEPS & DMRG & LSTM \\ \hline
 $-0.603535$ & $-0.6035218$ & $-0.603417$ \\ \hline
\end{tabular}
\end{center}
 \caption{Comparison between the ground state energy calculated using gradient optimized PEPS, DMRG keeping 4096 states, and our LSTM based Ansatz}
 \label{table:2dheisenberg}
\end{table}

While our model is able to achieve O($10^{-4}$) error, it is slightly outclassed by the best Monte Carlo and DMRG methods. However, the $6 \times 6$ lattice is suboptimal for our model, as edge effects are still extremely important. To examine how our model scales, we use iterative retraining to move from $6 \times 6$ lattice to a $30 \times 30$, adding two electrons to each edge at a time. Since $30 \times 30$ is larger than other lattices that have been simulated, we compare our energies with an extrapolation of the energies given in \cite{PEPSheisenberg}, which predicts the infinite size energy to very high accuracy.  

\begin{figure}[H]
\centering
  \includegraphics[width=.9\linewidth]{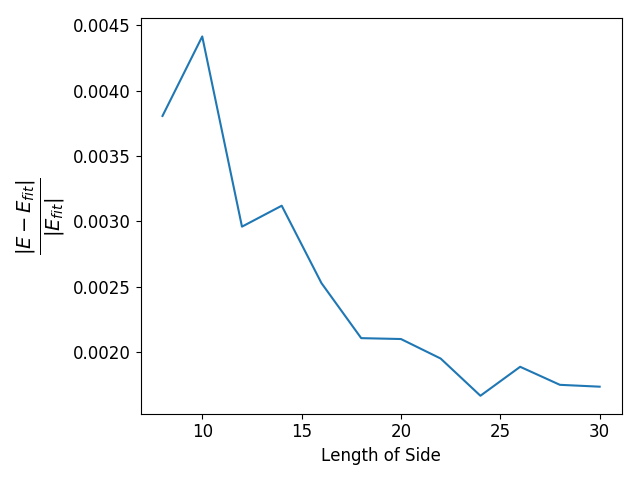}
  \caption{Error as a function of side length during the iterative retraining procedure. We compare our energies with those generated from an extrapolation given in \cite{PEPSheisenberg}}
      \label{fig:2dperformance}
\end{figure}

The performance is shown in figure \ref{fig:2dperformance}. Intriguingly, our performance improves as we scale to larger systems \footnote{The results on the $8 \times 8$ and $10 \times 10$ lattices may seem really poor. We'd like to emphasize that these are the results during the iterative retraining procedure where we only used $10^3$ minibatches and did not symmetrize our model. These models were stepping stones to move towards the regime where bulk effects are dominant}, affirming the notion that our RNNs are learning a model of the bulk. 

\section{Discussion}
We have introduced iterative retraining, a method for simulating entangled lattice models using recurrent neural networks. Unlike traditional Monte-Carlo methods, where the ground state is studied by simulating a small system, iterative retraining elucidates the bulk properties by scaling out edge effects. This is a useful paradigm shift, as macroscopic systems are dominated by the bulk. 

Iterative retraining can be used to simulate large models with tremendous data efficiency. Instead of requiring $O(10^6/10^7)$ samples from a large model (as is typically needed to train a model from scratch), iterative retraining requires a large number of samples from a small model and only a few samples from larger models (usually O($10^5$) and sometimes down to a few thousand for very large models). Since the speed of VMC scales polynomially in system size, iterative retraining saves a tremendous amount of computation for large systems. \footnote{In practice, even more time is saved, as the usual bottleneck is hyperparameter/architecture tuning. We found that training large models from scratch was often difficult due to local minima, whereas smaller models were much more straightforward to train.} 

Iterative retraining bears some similarities to density matrix normalization group (DMRG), where the system is grown by focusing on relevant regions of the Hilbert space. However, iterative retraining generalizes much better to higher dimensions, as the information flow can be shaped to respect the translational symmetries of the system. In contrast, DMRG models need to choose a single path through the system which is usually snake-like in 2D \cite{DMRG2D}. Furthermore, 2D generalizations of matrix product states, such as pair-entangled product states \cite{PEPSDescription} and isometric tensor states \cite{isotensor}, are usually entanglement limited. On the other hand, RNNs are comfortable representing highly entangled states \cite{RNNreconstruct}. Iterative retraining provides a roadmap for simulating 2D systems with arbitrary entanglement. 


We demonstrate that autoregressive models parameterized by RNNs provide a good Ansatz for iterative retraining. While we chose RNNs due to their ubiquity in machine learning, any translationally invariant model could be used. For example, one may want to model systems with long range order using a translationally invariant attention model \cite{allyouneed}, which tends to have longer memory. We believe that with more architecture and training scheme experimentation, iterative retraining could become the dominant method for understanding the bulk of strongly interacting systems.     

\section{\label{sec:appendix} Appendix}

\subsection{\label{sec:LSTM} LSTMs and GRUs}
The equations for an LSTM layer are given by:
\begin{eqnarray}
\vec{i}_t = \sigma(W_{is} \vec{S}_t + W_{ih} \vec{h}_{t-1} +  b_i), \\  
\vec{f}_t = \sigma(W_{fs} \vec{S}_t + W_{fh} \vec{h}_{t-1} + b_f), \\  
\vec{c}_t = \textrm{tanh}(W_{cs} \vec{S}_t + W_{ch} \vec{h}_{t-1} + b_c), \\  
\vec{o}_t = \sigma(W_{os} \vec{S}_t + W_{oh} \vec{h}_{t-1}  + b_o), \\
\vec{c}_t = \vec{i}_t*\vec{c}_t + \vec{f}_t*\vec{c}_{t-1}, \\
\vec{h}_t = \vec{o}_t*\textrm{tanh}(\vec{c}_t),
\end{eqnarray}
The LSTM maintains two hidden states, known as the hidden and cell states. The cell state, $\vec{c}_t$ is a persistant memory state, where information can only be erased using the forget gate $\vec{f}_t$. Information is read out through the hidden state $\vec{h}_t$ which controls the flow of short term information. The GRU is a much simpler model that also has persistant memory,
\begin{eqnarray}
\vec{i}_t = \textrm{tanh}(W_{is} \vec{S}_t + W_{ih} \vec{h}_{t-1} +  b_i), \\  
\vec{u}_t = \sigma(W_{us} \vec{S}_t + W_{uh} \vec{h}_{t-1} + b_u), \\  
\vec{h}_t = (1 - \vec{u}_t)*\vec{h}_{t-1} + \vec{u}_t*\vec{i}_t.
\end{eqnarray}
In GRUs, the update gate, $\vec{u}_t$ controls what is forgotten and written into memory. 

\subsection{Enforcing Symmetries}
We symmetrize our model in a similar manner to \cite{NAQS}, and \cite{LSTM}. We express the norm squared of the wavefunction for a basis state as an average of the norm squared of the wavefunction for the basis state transformed by all possible symmetry operations:
\begin{eqnarray}
|\psi(\vec{s})|^2 = \frac{1}{N_\tau} \sum_\tau |\psi(\tau \vec{s})|^2
\end{eqnarray}
We choose the phase to be the circulant mean of the phases of the transformed basis states, weighted by the norm squared of their wavefunctions.
\begin{eqnarray}
\phi(\vec{s}) = \textrm{Im}\Big[\textrm{log}\Big(\frac{\sum_\tau |\psi(\tau \vec{s})|^2 e^{i \phi(\tau \vec{s})}}{\sum_\tau |\psi(\tau \vec{s})|^2}\Big)\Big]
\end{eqnarray}
We can sample from the symmetrized model in the same manner as the unsymmetrized model, since the probability of sampling state $\vec{s}$ is the proportional to the probability of sampling any of the symmetry transformations, $\sum_\tau |\psi(\tau \vec{s})|^2$. 

\subsection{Hyperparameters and Regularization}
For all experiments in 1D we use the Adam optimizer \cite{Adam} and a minibatch size of 100 samples. For the 1D model we train our system of 80 spins with an initial learning rate of $\eta_0 = 10^{-3}$ decayed according to the function $\eta_t = \frac{\eta_0}{\sqrt{1 + 0.001*t}}$. We do iterative retraining with $\eta = 10^{-4}$. For all of our experiments we clip the $L2$ norm of our gradients to one. 

For our $6x6$ 2D model we first train an unsymmetrized model for $10^{4}$ minibatches with $\eta = 10^{-4}$. We then symmetrize our model and train for another $10^{4}$ minibatches with $\eta = 10^{-5}$. To fine tune our model, we increase the batch size to $1000$ and train for another $2000$ minibatches. 

For the iterative retraining procedure, we generalize from the unsymmetrized 6x6 model. At each iteration, we add two electrons to each edge and retrain starting from the checkpoint of the previous model. For $L = 8 \rightarrow 12$ we generalize using $1000$ minibatches at a learning rate of $\eta = 10^{-4}$. For $L = 12 \rightarrow 30$ we use $100$ minibatches with a learning rate of $10^{-5}$. We estimate the energy using the final $2500$ samples.   

We find that we are able to achieve more consistent performance by reshaping the optimization landscape. We add two terms to our loss function, a 'pseudo-entropy' reward that encourages our model to sample the Hilbert space more evenly, and a magnetization penalty, which moves our solution towards the spin-0 subspace. Our total cost function is:
\begin{eqnarray}
E + \frac{T}{N_e} \sum_{\vec{s} \sim |\psi(\vec{s})|^2} \textrm{log}(|\psi(\vec{s})|^2)  + C \sum_{\vec{s} \sim |\psi(\vec{s})|^2} M_{\vec{s}}^2. 
\end{eqnarray}
In the limit $T \rightarrow \infty$ the network tries to learn the phases in an equal amplitude configuration, which aligns with the finding that pretraining the phases improves performance \cite{signstruct}. For our models we took $T = 1/(1 + 0.001*\textrm{minibatch})$ and $C = 10$. During iterative retraining we took $T = 0$.

\section{Code}
Code for this project can be found at \url{https://github.com/chrisrothUT/Iterative-Retraining.git}

\section{Acknowledgements}
We'd like to thank Giuseppe Carleo for useful debugging tips and insight into optimizing parallelization. We'd also like to thank Allan MacDonald, Andrew Potter, and Chunli Huang for help editing the manuscript.  

While writing this manuscript, we became aware of another paper that also performed Variational Monte Carlo using RNNs \cite{RNNwf}. We came to our results independently, and are excited to see further verification that RNNs are an extremely useful tool for variational Monte-Carlo. We'd like to emphasize that the focus of our paper is iterative retraining, a method for learning the ground state wavefunction for bulk periodic systems.

\bibliographystyle{unsrt}
\bibliography{IterativeRetraining.bib}

\end{document}